# Real-time multimode dynamics of terahertz quantum cascade lasers via intracavity self-detection: observation of self mode-locked population pulsations

H. Li,[1] W. Wan,[1] Z. Li,[1] J. C. Cao,[1] S. Lepillet,[2] J-F. Lampin,[2] K. Froberger,[2] L. Columbo,[3,5] M. Brambilla,[4,5] and S. Barbieri.[2,*]

[1]*Key Laboratory of Terahertz Solid State Technology, Shanghai Institute of Microsystem and Information Technology, Chinese Academy of Sciences, 865 Changning road Shanghai, 200050, China*

[2]*Institute of Electronics, Microelectronics and Nanotechnology, University of Lille, ISEN, CNRS, UMR 8520, 59652 Villeneuve d'Ascq, France*

[3] *Dipartimento di Elettronica e Telecomunicazioni, Politecnico di Torino, Corso Duca degli Abruzzi 24, 10129 Torino, Italy*

[4] *Dipartimento di Fisica Interateneo, Università e Politecnico di Bari, Via Amendola 173, 70123 Bari, Italy*

[5] *IFN – CNR, Via Amendola 173, 70123 Bari, Italy*

*Corresponding author: stefano.barbieri@iemn.fr*

# Abstract

Mode-locking operation and multimode instabilities in Terahertz (THz) quantum cascade lasers (QCLs) have been intensively investigated during the last decade. These studies have unveiled a rich phenomenology, owing to the unique properties of these lasers, in particular their ultrafast gain medium. Thanks to this, in QCLs a modulation of the intracavity field intensity gives rise to a strong modulation of the population inversion, directly affecting the laser current. In this work we show that this property can be used to study the real-time dynamics of multimode THz QCLs, using a self-detection technique combined with a 60GHz real-time oscilloscope. To demonstrate the potential of this technique we investigate a free-running 4.2THz QCL, and observe a self-starting periodic modulation of the laser current, producing trains of regularly spaced, ~100ps-long pulses. Depending on the drive current we find two regimes of oscillation with dramatically different properties: a first regime at the fundamental repetition rate, characterised by large amplitude and phase noise, with coherence times of a few tens of periods; a much more regular second-harmonic-comb regime, with typical coherence times of ~$10^5$ oscillation periods. We interpret these measurements using a set of effective semiconductor Maxwell-Bloch equations that qualitatively reproduce the fundamental features of the laser dynamics, indicating that the observed carrier-density and optical pulses are in antiphase, and appear as a rather shallow modulation on top of a continuous wave background. Thanks to its simplicity and versatility, the demonstrated technique is a powerful tool for the study of ultrafast dynamics in THz QCLs.

# I. Introduction

QCLs operating in the mid-infrared and THz spectral regions have been shown experimentally to spontaneously generate frequency combs with various degrees of coherence [1-10]. In the mid-infrared (mid-IR) such combs present a predominantly quadratic phase dispersion (i.e. a constant group-delay) leading to a frequency modulated (FM) output, characterised by an approximately linear frequency chirp over one roundtrip period [11-15].

As pointed out in a number of works, a key ingredient leading to FM-type of mode-locking in QCLs is the fact that the gain recovery dynamics unfolds on a timescale much shorter than the roundtrip time. In this case, intensity oscillations of the intracavity field due to multimode beatings, generate a strong modulation of the population inversion [16, 17]. As shown in Refs. [13, 18] thanks to the fact that QCLs have a finite linewidth-enhancement factor (LEF), i.e. a sizeable phase-amplitude coupling, oscillations of the population inversion result into a strong non-linearity at all orders including the $3^d$ (Kerr-like). The latter, critically depending on the value of the group-velocity dispersion, can trigger a self-starting FM comb. If, on the one hand, the condition that the gain recovery time is much shorter than the roundtrip time is clearly satisfied in mid-IR QCLs, with their ~ps-long upper state lifetime, the situation is less clear-cut for THz QCLs. Reported gain recovery times for THz QCLs are rather in the 10-50ps [19-22] which appears to favour a more conventional amplitude-modulated (AM), or at least a mixed AM-FM type of mode-locking [23-25].

Regardless of the type of mode-locking mechanism and the type of the considered QCL lasers, the emission of self-starting QCL combs has been found to always present a sizeable amount of AM, in some cases, under the form of ps-long pulses with high contrasts [11, 13, 15, 26]. Such modulation is at the basis of so-called *beatnote spectroscopy* techniques [1], among which the *shifted wave interference Fourier transform* (SWIFT) *spectroscopy* is the most powerful [26, 27]. These techniques rely on the measurement of the difference frequency beatnote between adjacent longitudinal modes of the laser spectrum to assess their mutual coherence, and therefore determine the level of coherence of the entire comb spectrum.

Measuring the beatnote signal between adjacent modes, or fundamental beatnote, requires a quadratic detection process with a bandwidth (BW) at least of the order of the cavity free spectral range (FSR). Besides the use of a sufficiently fast external detector, such as a quantum well infrared photodetector [28] or a Schottky diode, it is well known that this function can be performed by the QCL itself simply by monitoring its current [29]. Indeed, as pointed out above, the intermodal optical beating generates, through partial gain saturation, a modulation of the population inversion $\Delta n$ [16, 17]. This, in turn, produces a modulation of the laser current, which, at first approximation, is directly proportional to $\Delta n$ [29, 30].

Since over two decades, intracavity self-detection of the fundamental beatnote has been routinely used for various applications, including studying the coherence of multimode THz QCLs [5, 29, 31]. However, it is clear that limiting the measurement to the fundamental oscillation at the FSR, somewhat undermines the potential of this technique. Indeed, the fact that the gain dynamics of THz QCLs unfolds on a time scale of few tens of ps, determined by intersubband relaxation, leads to self-detection BWs in the ~10-100GHz range [29, 32]. Hence, by monitoring the QCL current with fast electronic acquisition, one could observe *in real time* the temporal evolution of the intensity of the intracavity field propagating across the QCL resonator as a result of the coherent superposition of many lasing modes. To this end we note that although techniques such as SWIFT spectroscopy can indeed be used to reconstruct the time-dependent intensity of the laser field, nevertheless the resulting time-trace is inevitably averaged over tens of billions of periods.

Typically, the FSR of QCLs lies in the 10GHz, range (3-4mm cavity length), which, in terms of detection system BW requirements, poses serious challenges if one wants to observe the real-time evolution of a QCL determined by cavity modes separated by several FSRs. An attempt to observe in the time domain the field intensity emitted by an actively mode-locked 2.5THz QCL with a ~10GHz FSR was done by using a ~30GHz BW Schottky diode mixer, thus limiting the detection to the $3^d$ FRS harmonic [29]. To circumvent this limitation, here we have fabricated an exceptionally long THz QCL, based on a 15mm single-plasmon ridge waveguide, and relying on an ultra-low threshold, 4.2THz active-region design [33, 34]. The resulting multimode emission spectrum consists of ~40 longitudinal modes separated by a FSR of ~2.4GHz. The free-running QCL was driven in continuous wave (CW), and we monitored the *ac* current gain-induced modulation with the help of a microwave probe connected to a real-time oscilloscope, yielding an overall measurement-system BW of ~45GHz. This BW allowed us to probe in real time the mutual coherence of ~20 modes.

With this real-time acquisition technique, we perform a direct and complete analysis of the laser dynamics *vs* drive current using the collected time-dependent intensity waveforms (WFs, allowing to discriminate between amplitude and phase noise. Depending on the drive current, we observe a stable comb-regime characterised by two current pulses per roundtrip, alternating with wider current regions characterised by a highly unstable regime, with one pulse per roundtrip, and much more pronounced amplitude noise and timing jitter.

On the theoretical side we manage to reproduce a large part of the experimental results by integrating a set of Effective Semiconductor Maxwell-Bloch Equations already able to describe e.g. the alternation between locked and unlocked multimode regimes in a THz-QCL and the spontaneous formation of pulse trains in both the intracavity electric field intensity and carrier's density [18]. Notably, here we find the spontaneous generation of shallow optical and carrier's density pulses, on top of a CW background. These pulses, also designated as "pulsations" to distinguish them from traditional, high-contrast, mode-locked laser pulses, are consistent with the master equation approach presented in Ref. [13] and by the analytical results recently derived by D. Burghoff within an approximated mean field theory [14]. We believe that the demonstrated technique and results will pave the way to the study of the real time dynamics in THz QCLs.

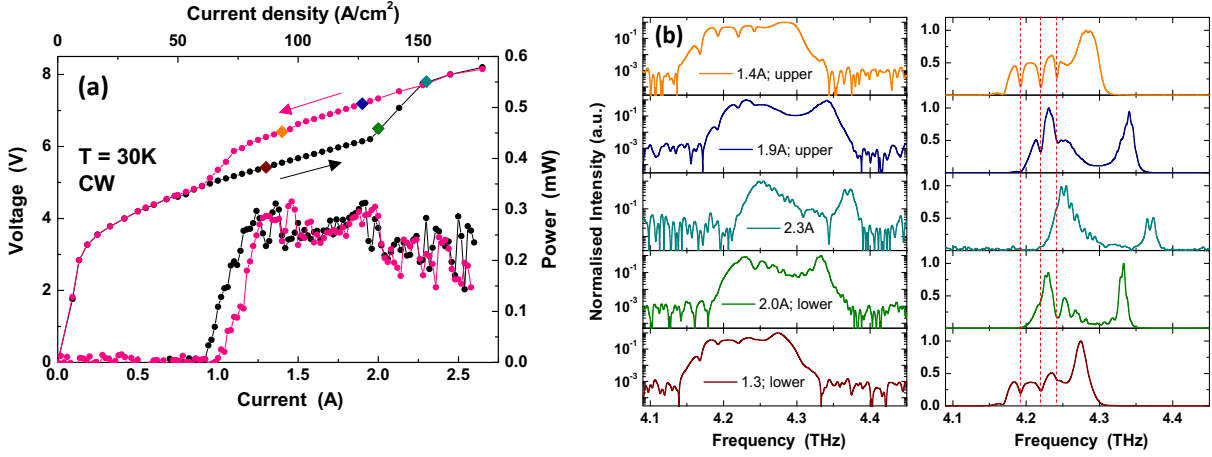

**Fig.1**. (a) V/I and P/I characteristics of the QCL under study. The laser was driven in CW at T = 30K. The drive current was swept continuously upward (black dots) and downward (pink dots). (b) Emission spectra measured with an FTIR spectrometer at different currents, corresponding to the colored dots of panel (a). Spectra are shown in logarithmic scale (left panel) and linear scale (right panel). The red dashed lines correspond to water absorption lines peaks.

## II. QCL *dc* characteristics and self-detection setup

The active region of the QCL studied in this work is based on 76 repeated periods of an $Al_{0.25}Ga_{0.75}As$-GaAs bound-to-continuum design, with a resonant-phonon extraction stage [32]. The precise layers sequence can be found in Supplementary material, and is a slight variation of the one reported in Ref.[35]. The heterostructure is grown on top of a semi-insulating GaAs substrate, and the active region is sandwiched between 400nm and 50nm thick bottom and top contact layers, *n*-doped at levels of $3 \times 10^{18}$ and $5 \times 10^{18}$ cm$^{-3}$ respectively. Devices were processed in 15mm-long, 100μm-wide single-plasmon waveguides.

In Fig.1(a) we report the *dc* Voltage/Current (V/I) and Ouput Power/Current (P/I) characteristics of the QCL studied here, measured at T=30K. These were obtained by a continuous up-sweep of the drive current from 0A to 2.65A, followed by a down-sweep from 2.65A to 0A. We observe a clear hysteretic behaviour in the V/I characteristic. As discussed in detail in Ref. [35] this is attributed to the presence of a space charge accumulation layer, progressively moving across the active region as the current is changed, and dividing the latter into a low and a high field-domain. Thanks to the very low threshold current density (60A/cm$^2$), the threshold current is of only 1A, despite the extremely long ridge-waveguide (THz QCL cavity lengths are typically of the order of only a few mm). The measured output power, of 0.2-0.3mW, is fairly constant, and we observe a very broad lasing current range extending up to 2.5 times the threshold and beyond. A few representative emission spectra, measured with an FTIR spectrometer, are reported in Fig.1(b). Emission covers up to ~200GHz, with the centre of the spectrum blue shifting from ~4.23THz to ~4.3THz from low to high current.

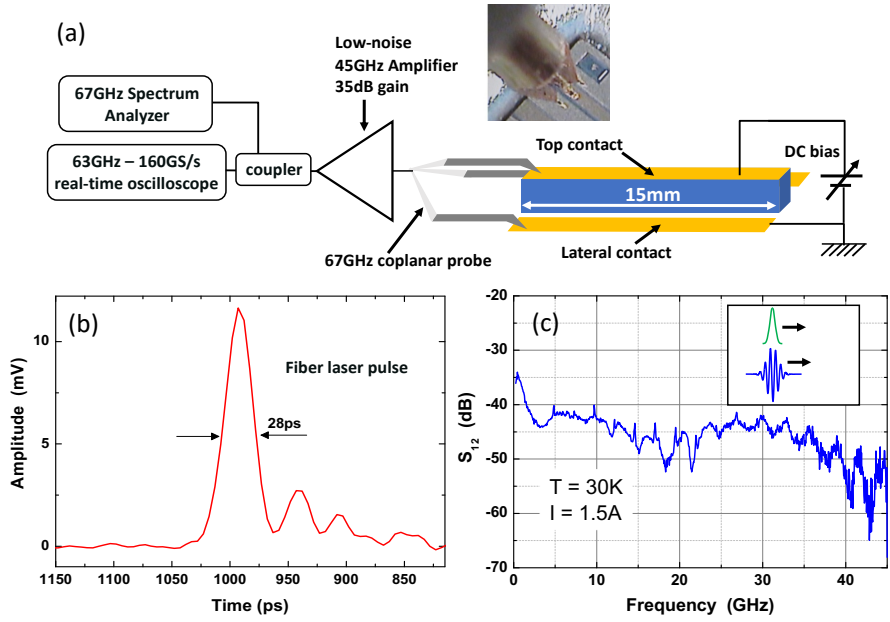

**Fig. 2**. (a) Schematic of the experimental setup. Inset. Microscope image of the microwave coplanar probe positioned at one end of the QCL ridge-waveguide. (b) Photocurrent pulse measured with the amplifier and real-time oscilloscope of panel (a). The pulse is obtained by illuminating an InGaAs PIN photodiode of ~30GHz 3dB-BW, with a train of ~150fs pulses from a mode-locked fiber-laser. (c) $S_{12}$ parameter of the QCL waveguide measured at 30K, under a drive current of 1.5A. Inset. Schematic drawing showing the intracavity THz pulse co-propagating with the current pulse in the resonator.

In Fig. 2(a), we report a schematic of the experimental setup used for the self-detection experiment. The QCL under study is mounted on the cold-plate of a cryogenic probe station and kept at a temperature of 30K. A broadband, 67GHz-BW coplanar probe of 125μm-pitch is positioned at one end of the QCL ridge cavity, at approximately 300μm from the facet. The coplanar probe is connected to a 45GHz-BW RF amplifier with a ~35dB gain (RF-Lambda RLNA00M45GA), whose output is monitored simultaneously with a 60GHz-BW real-time oscilloscope (Keysight Infiniium), and a 67GHz-BW spectrum analyser. The *dc*-bias is applied to the QCL using 4 standard *dc* probes: two probes are positioned on the top contact, and two probes on the lateral ground contacts (note that we obtained virtually the same data using a wire-bonded QCL, showing that the observed behaviour is independent from the type of electrical wiring).

To experimentally test the amplifier, we have connected it to a fast InGaAs photodiode with a ~30GHz ,3dB BW, that we illuminated with a train of ~150fs mode-locked pulses from a commercial fiber-laser. The resulting electrical pulse, measured with the real-time oscilloscope, is displayed in Fig2(b). Its ~28ps time-length is consistent with the response time of the photodiode, showing that our detection-system bandwidth is of at least ~35GHz.

In Fig.2(c) we report the $S_{12}$ parameter of the 15-mm long QCL ridge waveguide, measured with two coplanar probes at both ends of the device and a vector network analyser (the electric field profile of the microwave guided mode, computed with a finite element solver is shown in [Supplementary material](Supplementary material)). The QCL was operated at 30K, with a drive current of 1.5A. From 0 to 40GHz we observe a very strong power attenuation, between 40dB and 50dB, i.e. in the range ~2.5-3.3dB/mm, up to ~3.3-4dB/mm above 40GHz. This is attributed to the combined effect of the device conductance and free carrier absorption in the doped contact layers and metal contacts [36]. As a consequence of such a strong absorption, the propagation of the electrical pulse is heavily damped, preventing microwave resonant effects. In other words, as shown pictorially in the inset of Fig.2(c), the electrical pulse and the optical pulse co-propagate.

### III. Intra-cavity real-time self-detection

In Fig.4(a) and Fig.5(a) we report representative time-traces of WFs labelled WF-FN and WF-HM ("FN" for "fundamental and "HM" for harmonic, see below), measured using the real-time oscilloscope and the experimental setup shown in Fig.2 (see Supplementary material for additional WFs - note that as a result of the filtering of the amplifier the measured WFs have a zero *dc* component). They were recorded with a rising-edge trigger at *t* = 0 (with thresholds at 980mV and 408mV respectively), and by averaging 256 single-shot time-traces over an interval of 2ns. As shown by the pink arrows in Fig.3, the QCL was driven in CW at 6.6V-2A for WF-FN, and 7.6V-2.25A for WF-HM.

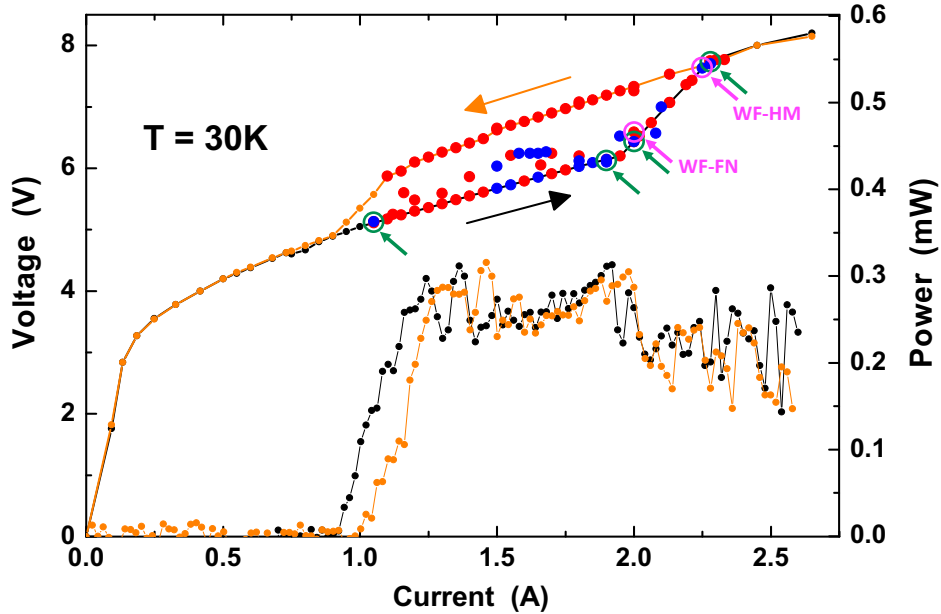

Fig.3. (a) V/I and Power/I characteristics of the QCL under study (black and orange dots – same as Fig.1(a)). The red and blue dots correspond to two different regimes, respectively with 1 and 2 pulses per roundtrip (see main text). The pink arrows indicate the two points of operation corresponding to the WFs, labelled FN and HM, displayed in Fig.4, Fig.5, and Fig.6. The green arrows indicate the operating points were both single and double pulse regimes could be triggered.

In the interval -1ns/+1ns, both WFs display a train of current pulses with a full width at half maximum (FWHM) of ~110ps and ~90ps for WF-FN and WF-HM respectively. These pulses are spontaneously generated by the QCL operating in free-running and are the result of the modulation of the population inversion by the intracavity field. They show that the QCL is operating in a regime of self-mode-locking, with a degree of temporal coherence that can be determined thanks to the real-time capability of our experimental technique.

A first striking feature is that while WF-FN shows one pulse per roundtrip, i.e. a repetition rate of 2.43GHz, corresponding to the cavity FSR, WF-HM shows two pulses per roundtrip, with a repetition rate at ~4.83GHz, at twice the FRS [23, 37]. Red and blue dots in Fig. 3 indicate the points on the QCL I/V curve with WFs showing respectively 1 or 2 pulses per roundtrip. These two regimes were triggered by randomly changing the QCL drive current upward or downward in a continuous way. As shown in the Figure, we observe multiple operating points, with the same drive current and different voltages, lying on top but even inside the hysteresis loop obtained by a full continuous up-sweep followed by a down-sweep (same as Fig.1(a)). In some cases (green arrows in Fig. 3), and within the error of the voltage measurement, the single and double pulsing regimes are superimposed, i.e. one regime or the other can be obtained at virtually the same bias and current, depending on the "current history".

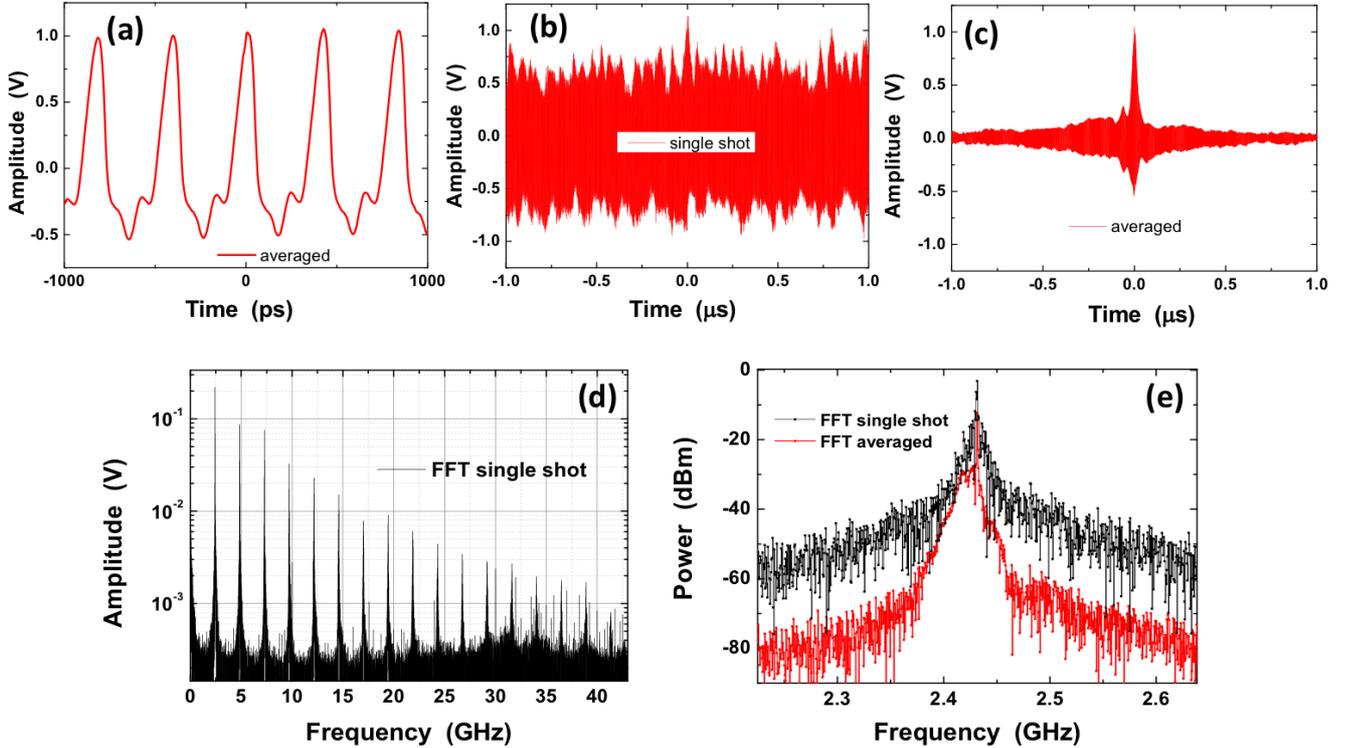

Fig. 4. WF-FN (a) Averaged time-trace in the interval -1ns/+1ns (256 averages). (b) Single-shot (no averaging) time-trace in the time interval -1μs/+1μs, corresponding to ~5000 periods (see Movie 1 for the video). (c) Averaged time-trace in the time interval -1μs/+1μs (256 averages). (d) Fourier transform of the time-trace of panel (b). (e) Close-ups on the fundamental comb line (2.43GHz) of the spectrum of panel (d) (black) and of the Fourier transform (not shown) of the time-trace of panel (c) (red).

Thanks to real-time acquisition, in Fig.4, Fig.5 and Fig.6, we provide a complete analysis of the measured WFs, revealing two completely different regimes.

Fig.4(b) shows a time-trace of WF-FN recorded in a single-shot (no averaging) in the time interval -1μs/+1μs, corresponding to ~5000 periods. We observe a strong quasi-random modulation of the signal amplitude, or amplitude noise. In addition, the shape of the current pulses dramatically changes with time. This is shown in the panels of Fig.6(a), displaying three ~2.5ns-long time windows of the time-trace of Fig.4(b), at different delays from the trigger event (~-1μs, ~0μs or~+1μs). An alternative view of the extent of such phenomenon can be found in Fig.6(b) and Fig.6(c), showing WF-FN recorded for two different values of the trigger level, and with the oscilloscope set in *persistence* mode. In this way 2ns-long time-traces are recorded in sequence and shown simultaneously on the display: more frequent WFs appear with a brighter colour intensity.

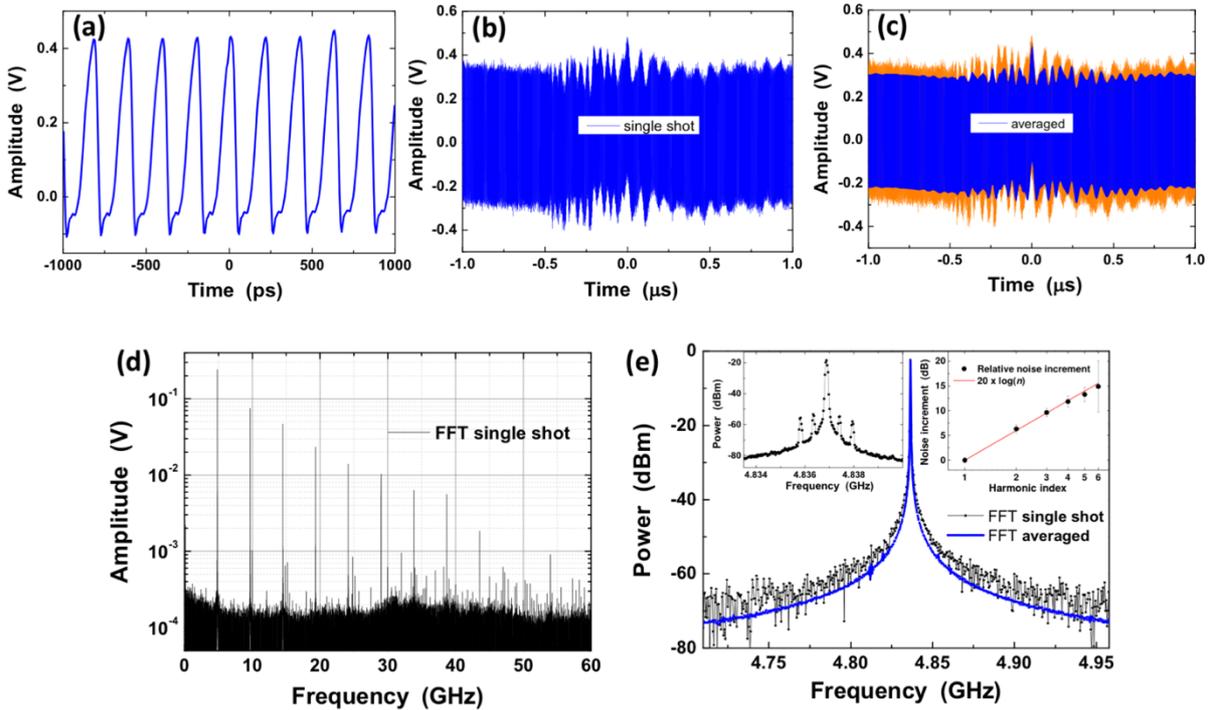

Fig. 5. WF-HM (a) Averaged time-trace in the interval -1ns/+1ns (256 averages). (b) Single-shot (no averaging) time-trace in the time interval -1μs/+1μs, corresponding to ~5000 periods (see Movie 2 for the video). (c) Blue curve: averaged time-trace in the time interval -1μs/+1μs (256 averages). Orange curve: same time-trace of panel (b). (d) Fourier transform of the time-trace of panel (b). (e) Close-ups on the fundamental comb line (4.83GHz) of the spectrum of panel (d) (black) and of the Fourier transform of the time-trace of panel (c) (blue). Left inset. Power spectrum of the fundamental comb line measured with the spectrum analyser shown in Fig.2. Right inset: noise increment of the different harmonics with respect to the fundamental comb line, extracted from the Fourier transform of the averaged time-trace of panel (c). The noise was measured at 130MHz from each carrier (see the spectra in Supplementary material). The solid red line represents the $n^2$ dependence.

As can be seen, for a given trigger level, despite keeping a relatively regular periodicity, the amplitude and shape of the pulses fluctuate wildly. Changing the level of the trigger produces an even larger change (compare Fig.6(b) and Fig.6(c)), revealing the quasi-chaotic nature of the radiation emitted by the QCL (see also Movie 1). We systematically observed such type of behaviour when the QCL was operated in correspondence to the red dots of Fig.3, i.e. when emitting one pulse per roundtrip (see Supplementary material for more traces). The Fourier transform of the time-trace of Fig.4(b) is displayed in Fig.4(d). As expected we find a comb of lines separated by the cavity FSR, resulting from the coherent beating of the THz modes. From their number, we conclude that modes separated by up to at least ~17 FSRs contribute to the observed current pulse.

In Fig.4(c) we report the time-trace of WF-FN obtained by averaging over 256 time-traces in the interval -1μs/+1μs (we verified that by increasing further the number of averages the time-trace did not change). Contrary to the single-shot acquisition of Fig.4(b), in this case the amplitude of the signal undergoes a fast decrease with time. This results from the averaging process: as the pulse train loses its coherence, due to the strong amplitude noise and pulse shape fluctuations, the time-trace averages out. Within ~20ns from the trigger event (~50 periods) the pulse amplitude has decreased by a factor 2. In the frequency domain such loss of coherence gives rise to a large noise pedestal around the comb lines, of approximately 20MHz FWHM (see Fig.4(e) for the fundamental comb line).

The picture changes completely in the case of two pulses per roundtrip (WF-HM and blue dots in Fig.3). As shown in Fig.5(b) and Fig.5(c), in this case the pulse amplitude is virtually unaffected by the averaging process, and the fast decrease found in Fig.4(c) has disappeared. Amplitude fluctuations, whose magnitude is reduced compared to Fig.4(b), are mostly due to a coherent modulation of unknown origin (clearly visible in the averaged time-trace shown in Fig.5(c)). As shown by the three panels of Fig.6(d) and by the persistence-mode acquisition shown in Fig.6(e), the strong pulse-shape fluctuations found on WF-FN (Fig.6(a)) have completely disappeared (see also Movie 2). We systematically observed such type of behaviour when the QCL was operated in correspondence to the blue dots of Fig.3 (see Supplementary material for more traces). Overall, this results into a comb spectrum with orders of magnitude narrower lines (compare Fig.4(d) and Fig.5(d)), reflecting a much longer coherence time. In Fig. 5(e) we report a close-up of the Fourier transforms of the single-shot and averaged time-traces (Fig.5(b) and Fig.5(c)) around the first comb-line at ~4.8GHz. The lines are virtually identical and narrow, with no sign of the noise pedestal found in Fig.4(e). In the left inset we report a spectrum of the same line acquired with the spectrum analyser (see Fig.3(a)). The linewidth of ~50kHz is limited by the resolution bandwidth (30kHz), indicating a coherence time of the pulse train > 30μs.

To verify that WF-HM is dominated by timing jitter, i.e. that the related comb spectrum is dominated by phase-noise, we have measured the noise at ~130MHz from each comb line in the spectrum obtained by Fourier transforming the averaged time-trace (see the spectra in Supplementary material). The result is shown in the right inset of Fig.5(e), where we report the noise increment in dB with respect to the noise of the fundamental comb line at 4.83GHz *vs* the harmonic index $n$, where $n$ = 1, 2, 3… (the frequency of each comb line is given by $f_n = n \times 4.83\text{GHz}$). As shown by the red line, within the experimental error, the points follow the expected $20 \times \log(n)$ dependence [38].

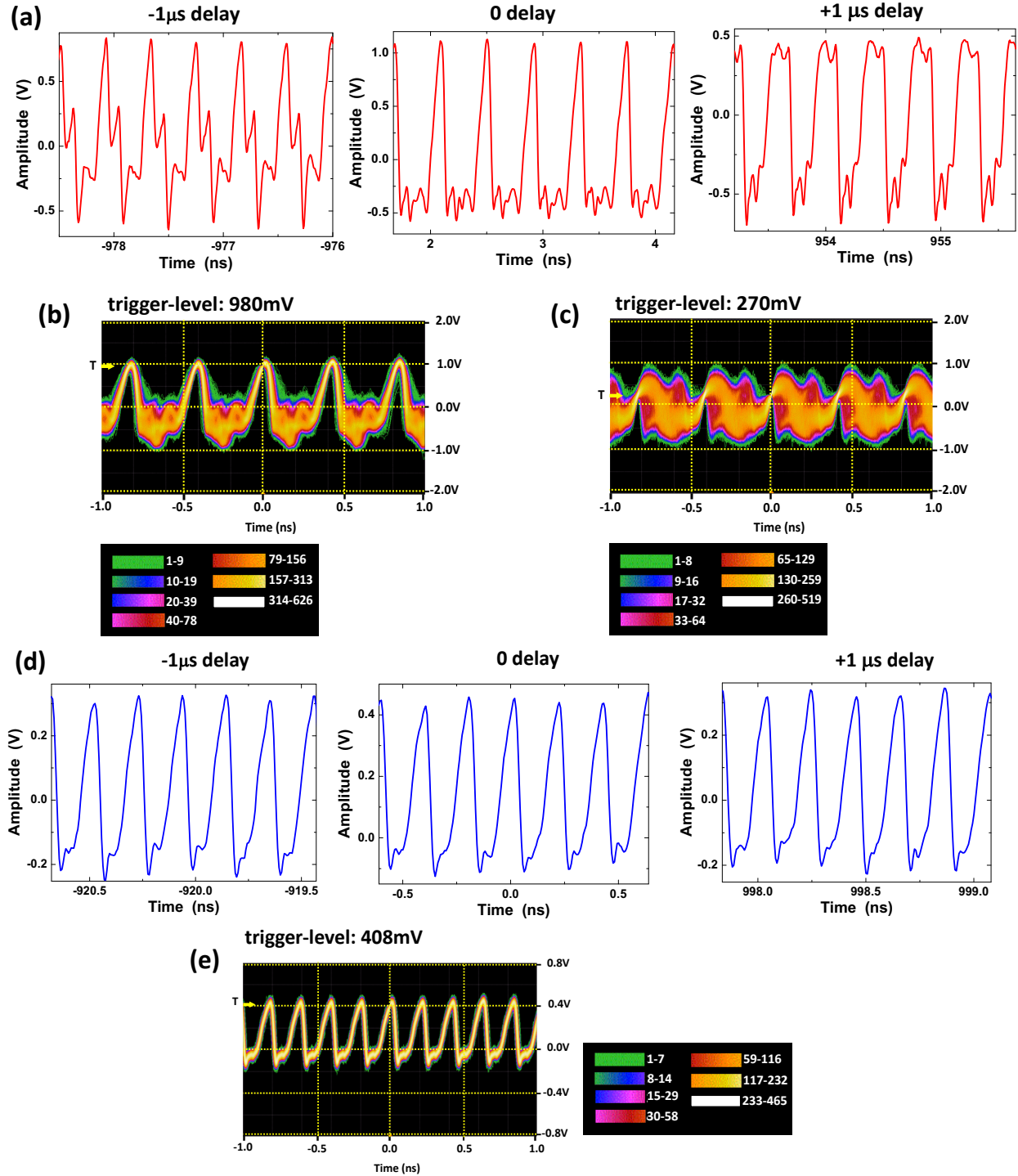

Fig. 6. (a) WF-FN. Time windows obtained from the time-trace of Fig.4(b), at different delays from the trigger event (~ -1µs, ~0µs or ~ +1µs). (b) WF-FN recorded with the oscilloscope set in *persistence* mode, with a trigger level of 980mV (same as for panel (a)). Single shot time-traces are recorded sequentially and superimposed on the display of the oscilloscope. More frequent FWs appear with a brighter colour intensity (c) WF-FN recorded with the oscilloscope set in *persistence* mode, with a trigger level of 270mV. (d) WF-HM. Time windows obtained from the time-trace of Fig.5(b), at different delays from the trigger event (~-1µs, ~0µs or~+1µs). (e) WF-HM recorded with the oscilloscope set in *persistence* mode, with a trigger level of 408mV (same as for panel (d)).

An important finding of our measurements is that the duration of the current pulses is always in the range ~90-110ps, regardless of the operating regime (single or double pulse per roundtrip). Following Fig.2(b) we exclude that this duration is affected by the measurement system bandwidth. At the same time the observed pulse length is somewhat longer than what would be expected from the THz emission spectra of Fig. 1 in the case of negligible or low phase dispersion, which points towards an FM-type mode-locking operation (see next Section).

### IV. Simulations and discussion

The experimental results reported in the previous Section could be reproduced with a good agreement by means of simulations based on a model developed in [18], and already successful in describing the multimode dynamics in a THz QCL based on a set of Effective Semiconductor Maxell Bloch Equations. With this approach we properly take into account peculiar features of the radiation-matter interaction in a semiconductor active medium, such as, most importantly, a non-null phase-amplitude coupling provided by the LEF. Its fundamental role in affecting both the stability of the continuous wave solution (single mode operation) and thus the multi-mode competition emerging thereof, as well as the self-phase locking leading to comb formation in a QCL has been very recently demonstrated [39].

| $\tau_d$ | $\tau_e$ | $\tau_p$ | $l$ | $n$ | $\alpha$ | $\Gamma$ |
|---|---|---|---|---|---|---|
| 0.1ps | 10ps | 500ps | 30mm | 3.3 | 1.2 | 0.032 |

Table 1. Values of the physical parameters used in the simulations.

Remarkably, in agreement with what demonstrated in Ref. [18] we show that the unidirectional ring configuration considered by our model is capable of reproducing the fundamental phenomenology of the experimental findings, although it does not encompass Spatial Hole Burning (SHB) associated to the superposition of forward and backward propagating fields in a FP resonator. This allows for a radical numerical simplification and potentially allows for a simplified analysis of the self-mode locking phenomenon.

The set of nonlinear partial differential equations that describe in a semiconductor laser, and thus in particular in a QCL, the spatio-temporal evolution of the slowly varying envelopes of the electric field $E$, the macroscopic polarization $P$ and of the carrier density $D$ in the rotating wave and low transmissivity approximations can be cast in the form:

$$\frac{c\tau_d}{nl}\frac{\partial E}{\partial z} + \frac{\partial E}{\partial t} = \sigma(-E + P) \quad (1)$$

$$\frac{\partial P_F}{\partial t} = \Gamma(1-i\alpha)[-P + (1-i\alpha)DE] \quad (2)$$

$$\frac{\partial D}{\partial t} = b\left[\mu - D - \frac{1}{2}\{E^*P + EP^*\}\right] \quad (3)$$

The dynamical variables $E, P, D$ are suitably normalized as specified in Ref. [18], and periodic boundary conditions are considered: the time $t$ is scaled on the shortest system time scale represented by the dipole dephasing time $\tau_d$ while the longitudinal spatial variable $z$ is scaled to the cavity length $l$ (equal to twice the experimental QCL ridge length). The other physical and geometrical parameters characteristic of the considered configuration are defined as follows: $\sigma = \frac{\tau_d}{\tau_p}$, and $b = \frac{\tau_d}{\tau_e}$ are the ratios of the dipole dephasing time to the photon ($\tau_p$) and to the carriers ($\tau_e$) life time respectively; $v = \frac{c}{n}$ is the group velocity of the light in the medium; $\alpha$ is the LEF; $\Gamma/\pi\tau_d$ is a measure of the FWHM of the gain/dispersion curve; $\mu = \frac{I}{I_{th}}$ is the ratio between the bias current and its value at threshold.

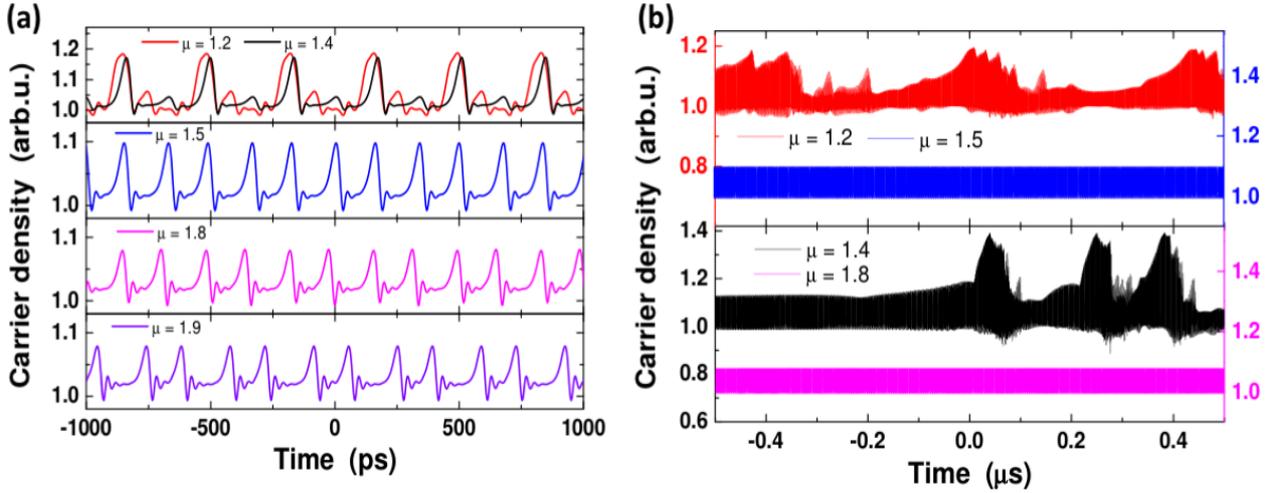

Fig.7 (a) Simulated carrier density WFs in the interval -1ns/+1ns showing the propagation of short pulses in the cavity for different values of the bias current normalized to the threshold current. Top panel: one pulse per roundtrip. Remaining panels: two pulses per roundtrip. (b) Carrier density WFs in the interval -0.5μs/+0.5μs corresponding to one and two pulses per roundtrip. Colors are as in panel (a). WFs with one pulse per roundtrip show strong amplitude fluctuations. On the contrary the 2-pulses per roundtrip temporal traces show a high degree of temporal coherence. The parameters used in the simulations are those reported in Table 1.

All simulations were performed adopting the parameter values reported in Table 1 which are fully compatible with the experimental configuration except for the FWHM of the gain lineshape ($\Gamma/\pi\tau_d \cong$ 100GHz) that is smaller than the typical expected values of ~400-500GHz (see below) [33,34].

In Fig. 7 we report the results of the numerical integration of Eqs. (1)-(3) done by using a split step method based on a pseudospectral and a Runge Kutta algorithms. In particular Fig. 7(a) shows the time-traces of the carrier density corresponding to steady state regimes with 1-pulse or 2-pulses per roundtrips respectively, and for different values of the parameter $\mu = \frac{I}{I_{th}}$. Figure 7(b) shows the same dynamical regimes on a much longer time scale from which it appears that only the solutions with 2-pulses are characterized by a low amplitude noise. In terms of pulse shape, these results are in good qualitative agreement with the time-traces reported in Fig. 4, 5, 6 and described in detail in the previous Section. Moreover, unstable 1-pulse solutions are systematically obtained for values of $\mu \lesssim 1.5$ while the stable solutions with 2-pulses are obtained for higher values of $\mu$. This is in fair agreement with the experimental results, as can be seen from Fig.3, where, in the lower branch of the IV curve, unstable single-pulse WFs appear predominantly for currents below 1.5A.

Concerning the pulse lengths (FWHM), regardless of the operating regime (stable or unstable) we obtain values in the range ~50-70ps (see Fig.7), i.e. shorter than what found experimentally (~90-110ps, see previous Section). This parameter depends on the value of the carrier's lifetime $\tau_e$. We found that to reach the observed ~100ps pulse length, the value of $\tau_e$ must be of about 100ps, which is nevertheless larger than the typical gain recovery times measured in the literature, in the range ~10ps- 50ps [19- 22].

From Fig.7 it emerges that the pulses appear as a shallow modulation ($\lesssim 20\%$), or pulsation [14], of the total carrier density, i.e. the optical power is dominated by a large continuous wave (CW) background. From Fig.8 we also find that the optical and carrier density pulses are $\cong$180deg out of phase. This finding and the observation of shallow pulses sitting on a CW background are expected since $\tau_e$ is much shorter than the roundtrip time, which has been shown to hamper the generation of strong pulses as in standard mode-locking, and rather lead to a CW dominated, FM-comb type of emission [16, 17]. We note that in the actual self-detection configuration (see Fig.2) it is clearly not possible to establish experimentally the presence of a CW background. Indeed such background gives rise to a *dc* voltage offset that is indistinguishable from the externally applied *dc* bias.

As discussed previously, the results of Fig.7 and Fig.8 are obtained with a gain FWHM of 100GHz. Indeed, while for this value we obtain a good agreement with the experimental results in terms of (i) the extension of the current regions showing stable/unstable pulsed emission, (ii) the number of pulses per roundtrip and (iii) the pulse shape, we found that an increased gain FWHM in the range from 200 to 400GHz (more in agreement with the experimental data [31]), while leading to the formation of temporal structures of similar shape, produced a regime with more than two localized pulses per roundtrip. Further increasing the FWHM to 500GHz showed the onset of either CW emission or of an irregular multimode regime, flagging a higher mode competition. We argue that this discrepancy might be ascribed to the unidirectional character of the field propagation in our model. In fact, in FP resonators, bidirectional propagation with inclusion of SHB evidenced modified single-mode instability and mode coupling, favoring mode-locking with a large gain FWHM, as recently shown in a mid-IR QCL [13-15]. Nevertheless, we expect that the qualitative dynamical picture should not be radically modified, showing a first transition from single-mode emission to multimode regimes, with alternating windows of localized pulses and unlocked, irregular emission [15, 18]. We plan to expand our model in the near future to include a FP configuration, following what done in Ref. [15] in the mid-IR, to achieve a more quantitative agreement with the experimental results presented here.

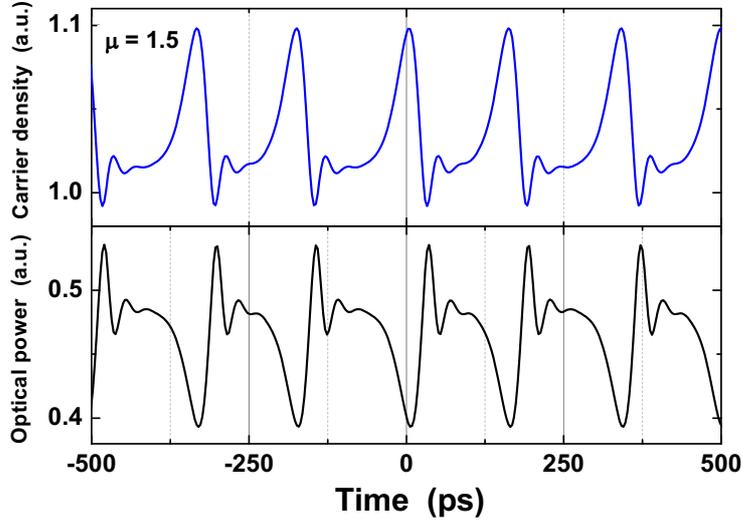

Fig.8 Simulated carrier density (blue - same as Fig.7), and associated intracavity optical power (black) in the interval -0.5ns/+0.5ns, for $\mu = 1.5$.

A closer inspection of Fig. 7(a) in the case of two-pulses per roundtrip ($\mu = 1.5, 1.8, 1.9$), reveals that the two pulses are not equally spaced in time, but rather appear as doublets separated by the roundtrip time. This fact is in disagreement with the experiment, where the pulses have a period equal to half the roundtrip time. The effect is progressively more pronounced when the value of $\mu$ is increased (see the bottom WF in Fig.7(a)). While we do not have an intuitive explanation for this phenomenon, we note that the solitonic character of the peaks, theoretically predicted in Refs. [14, 18], may lead, at steady-state, to an adjustment of the separation between localized structures arising from mode-locking. In the experiment, additional confinement mechanisms, not encompassed in the model, may be at work. As a result, and contrary to the experimental RF spectra, where only the lines with frequencies equal to even multiples of the cavity FSR are visible (see Fig.5(d) and Supplementary material), in the computed RF spectra, both even and odd multiples are present, with an attenuation of ~20 dB between the intensities of the first and second beatnote at $\mu = 1.5$ (see Supplementary material).

A final remarkable evidence provided by the model is the confirmation of coexisting regimes of 1-pulse and 2-pulse emission for the same parameters (bias current, in particular). Simulations show that either regime may appear at steady state, by using different random and noisy initial conditions. This exactly matches the experimental evidence, where the operating points showing coexisting regimes are reported by the green arrows in Fig.3 and are obtained by randomly ramping the current up and down, as described in the previous Section. We believe that this is a preliminary indication of the basic solitonic characteristics of the pulses emerging from the mode-locking onset, as already highlighted in Refs. [14, 18, 39]. Here, in addition, we can directly observe the carrier's density shape in the time domain and correlate it with our simulations. To the best of our knowledge, this is the first time that such a coupled experimental/numerical observation is provided in a THz-QCL.

## V. Conclusions

Understanding in depth the temporal dynamics of THz QCLs is proving crucial for the development of frequency combs based on these unique laser sources. To this end, the possibility to observe their behavior unfolding in real-time on timescales of tens of ps or lower, may bring to new insights. In this work we make an important step in this direction by demonstrating a self-detection technique based on broad BW electronic acquisition. So far, the scarcity of versatile, sensitive, and sufficiently fast THz detectors has hampered real-time measurements. On the contrary our technique does not employ an external detector, and can therefore be applied to any THz QCL, regardless of its emission wavelength, using the same experimental setup, which makes it an extremely appealing and flexible tool.

To demonstrate the potential of this technique we investigate a multimode THz QCL with a 15mm-long ridge cavity. By operating the QCL in free-running, we observe the spontaneous generation of ~100ps-long (not limited by the electronic system's bandwidth), regularly spaced current pulses propagating back and forth inside the cavity. The latter are the result of a modulation of the laser population inversion by the intensity of the QCL intracavity field, and are produced by the coherent interference of ~20 FP modes. As such they are the manifestation of so-called dynamic population gratings, or population pulsations [16, 31, 40, 41], whose spatiotemporal evolution was never observed in real time.

Depending on the pump current we find two clearly distinct dynamical regimes of operation. A first dominant regime shows one pulse per cavity roundtrip, and is characterized by high level of amplitude and phase noise, yielding coherence times of tens of ns. A second regime is instead observed predominantly at higher currents on a more limited current range, and shows two (evenly-spaced) pulses per roundtrip. In this case we find a stable pulse train, characterized by a much lower amplitude and phase noise, with coherence times of tens of μs.

The observation of regular and irregular multimode regime in different current regions, yielding respectively narrow and broad beatnotes at the cavity FSR is typical of multimode QCLs [1-10, 18]. In this work we have provided for the first time a real-time analysis of such regimes. Our experimental findings are corroborated by simulations obtained using a set of Effective Semiconductor Maxwell-Bloch Equations that includes a non-null phase-amplitude coupling provided by the LEF [18]. Despite the fact that our simplified model does not include SHB, we managed to reproduce the most salient experimental finding, namely the appearance of unstable/stable operating regions with one or two pulses per roundtrip respectively. The simulated carrier density (i.e. current) pulses are found to be 180deg out of phase [17] with respect to the intracavity optical pulses, and the QCLs shows a predominantly FM-type dynamics, with both (optical and carrier density) pulses appearing as a shallow modulation on top of a CW background. Importantly, our model confirms the coexistence of regimes with one or two pulses per roundtrip found experimentally at specific bias points (Fig. 3, green arrows), which we interpret as an indication pointing towards the solitonic character of the observed QCL dynamics. Indeed, a definite proof of this character would require further simulations, e.g. showing that the pulses can be excited independently within the same parameter choice by appropriately tailoring the initial conditions, or injecting pulses via a coherent external field. Further work will be aimed at elucidating these properties.

**Supplementary Material.** See supplementary material for: (i) the QCL active region layer's sequence and fabrication details; (ii) the computed electric field map of the microwave mode; (iii) the spectra of the different harmonics extracted from the Fourier transforms of the time traces of Fig.4 and Fig.5; (iv) additional waveforms; (v) the Fourier transforms of the simulated time traces of Fig.7.

**Acknowledgements.** We thank Giorgio Santarelli for very helpful discussions and suggestions, Pascal Szriftgiser for lending the ultrafast oscilloscope, and Guillaume Ducournau for assistance with the use of the broadband amplifier. This work was partially supported by Nord-Pas de Calais Regional Council; Fonds Européens de Développement Régional. RENATECH (French Network of Major Technology Centres). CPER "Photonics for Society". National Natural Science Foundation of China (61875220, 62035005, 61927813, and 61991432); "From 0 to 1" Innovation Program of the Chinese Academy of Sciences (ZDBS-LY-JSC009); National Science Fund for Excellent Young Scholars (62022084).

**Conflict of interest.** The authors have no conflicts to disclose.

**Data availability.** The data that support the findings of this study are available from the corresponding author upon reasonable request.

**References**

[1] A. Hugi, G. Villares, S. Blaser, H. C. Liu, and J. Faist, Nature **492**, 229–233 (2012).

[2] D. Burghoff, T.-Y. Kao, N. Han, C. W. I. Chan, X. Cai, Y. Yang, D. J. Hayton, J.-R. Gao, J. L. Reno, and Q. Hu, Nature Photon. **8**, 462–467 (2014).

[3] M. Rösch, G. Scalari, M. Beck, and J. Faist, Nature Photon. **9**, 42–47 (2015).

[4] H. Li, P. Laffaille, D. Gacemi, M. Apfel, C. Sirtori, J. Leonardon, G. Santarelli, M. Rösch, G. Scalari, M. Beck, J. Faist, W. Hänsel, R. Holzwarth, and S. Barbieri, Opt. Express **23**(26), 33270–33294 (2015).

[5] M. Wienold, B. Röben, L. Schrottke, and H. T. Grahn, Opt. Expr. **22**, 30410–30424 (2019).

[6] G. Villares, S. Riedi, J. Wolf, D. Kazakov, M. J. Süess, P. Jouy, M. Beck, and J. Faist, Optica **3**, 252–258 (2016).

[7] D. Kazakov et al. M. Piccardo, Y. Wang, P. Chevalier, T. S. Mansuripur, F. Xie, C. Zah, K. Lascola, A. Belyanin and F. Capasso, Nat. Photonics **11**, 789–792 (2017).

[8] K. Garrasi, F. P. Mezzapesa, L. Salemi, L. Li, L. Consolino, S. Bartalini, P. De Natale, A. G. Davies, E. H. Linfield, and M. S. Vitiello, ACS Photonics **6**, 73–78 (2019).

[9] A. Forrer, M. Franckie, D. Stark, T. Olariu, M. Beck, J. Faist, and G. Scalari, ACS Photonics **7**, 784−791 (2020)

[10] M. Jaidl, N. Opacak, M. A. Kainz, S. Shonhuber, D. Theiner, B. Limbacher, M. Beiser, M. Giparakis, A. M. Andrews, G. Strasser, B. Schwartz, J. Darmo, K. Unterrainer, Optica **8**, 780 (2021).

[11] M. Singleton, P. Jouy, M. Beck, and J. Faist, Optica **5**, 948–953 (2018).

[12] J. Hillbrand, A. M. Andrews, H. Detz, G. Strasser, and B. Schwarz, Nat. Photonics **13**, 101 (2019).

[13] N. Opačak, and B. Schwarz, Phys. Rev. Lett. **123**, 243902 (2019).

[14] D. Burghoff, Optica **7**, 1781–1787 (2020).

[15] C. Silvestri, L. L. Columbo, M. Brambilla, and M. Gioannini, Opt. Express **28**, 23846–23861 (2020).

[16] T. S. Mansuripur, C. Vernet, P. Chevalier, G. Aoust, B. Schwarz, F. Xie, C. Caneau, K. Lascola, C.-E. Zah, D. P. Caffey, T. Day, L. J. Missaggia, M. K. Connors, C. A. Wang, A. Belyanin, and F. Capasso, Phys. Rev. A **94**, 063807 (2016).

[17] M. Piccardo, P. Chevalier, B. Schwarz, D. Kazakov, Y. Wang, A. Belyanin, and F. Capasso, Phys. Rev. Lett. **122**, 253901 (2019).

[18] L. L. Columbo, S. Barbieri, C. Sirtori, and M. Brambilla, Opt. Express **26**(3), 2829–2847 (2018)

[19] R. P. Green, A. Tredicucci, N. Q. Vinh, B. Murdin, C. Pidgeon, H. E. Beere, and D. A. Ritchie, Phys. Rev. B **80**, 075303 (2009).

[20] J. R. Freeman, J. Maysonnave, S. Khanna, E. H. Linfield, A. G. Davies, S. S. Dhillon, and J. Tignon, Phys. Rev. A **87**, 063817 (2013).

[21] D. R. Bacon, J. R. Freeman, R. A. Mohandas, L. Li, E. H. Linfield, A. G. Davies, and P. Dean, Appl. Phys. Lett. **108**, 081104 (2016).

[22] C. G. Derntl, G. Scalari, D. Bachmann, M. Beck, J. Faist, K. Unterrainer, and J. Darmo, Appl. Phys Lett. **28**, 181102 (2020).

[23] A. Mottaghizadeh, D. Gacemi, P. Laffaille, H. Li, M. Amanti, C. Sirtori, G. Santarelli, W. Hansel, R. Holzwart, L. H. Li, E. H. Linfield, and S. Barbieri Optica **4**, 170 (2017).

[24] P. Tzenov, D. Burghoff, Q. Hu, and C. Jirauschek, Opt. Expr. **24**(20), 23232–23247 (2016).

[25] F. Cappelli, L. Consolino, G. Campo, I. Galli, D. Mazzotti, A. Campa, M. S. de Cumis, P. C. Pastor, R. Eramo, M. Rösch, M. Beck, G. Scalari, J. Faist, P. D. Natale, and S. Bartalini, Nature Photon. **13**, 562–568 (2019).

[26] D. Burghoff, Y. Yang, D. J. Hayton, J.-R. Gao, J. L. Reno, and Q. Hu, Opt. Expr. **23**, 1190–1202 (2015).

[27] Z. Han, D. Ren, and D. Burghoff, Opt. Expr. **28**, 6002–6017 (2020).

[28] M. Hakl, Q. Lin, S. Lepillet, M. Billet, J. Lampin, S. Pirotta, R. Colombelli, W. Wan, J. C. Cao, H. Li, E. Peytavit, and S. Barbieri, ACS Photon. **8**, 464–471 (2021).

[29] P.Gellie, S.Barbieri, J.F.Lampin,P.Filloux,C.Manquest,C.Sirtori,I.Sagnes,S.P.Khanna,E.H.Linfield, A. G. Davies, H. Beere, and D. Ritchie, Opt. Expr. **18**, 20799-20816 (2010).

[30] J. Faist in Quantum Cascade Lasers 112-124 (Oxford, 2013)

[31] M. Piccardo, D. Kazakov, N. A. Rubin, P. Chevalier, Y. Wang, F. Xie, K. Lascola, A. Belyanin, and F. Capasso, Optica **5**, 475–478 (2018)

[32] A. Forrer, Y. Wang, M. Beck, A. Belyanin, J. Faist, and G. Scalari, Self-starting Harmonic Comb Emission in THz Quantum Cascade Lasers, arXiv:2102.03033 [physics.optics] (2021)

[33] W. J. Wan, H. Li, T. Zhou, and J. C. Cao, Sci. Rep. **7**, 44109 (2017).

[34] Z. Li, W. Wan, K. Zhou, X. Liao, S. Yang, Z. Fu, J. C. Cao, and H. Li, Phys. Rev. Appl. **12**(4), 044068 (2019).

[35] M. Wienold, L. Schrottke, M. Giehler, R. Hey, and H.T. Grahn, J. Appl. Phys. **109**, 073112 (2011)

[36] W. Maineult, L. Ding, P. Gellie, P. Filloux, C. Sirtori, S. Barbieri, J−F. Lampin, T. Akalin, H. E. Beere, and D. A. Ritchie, Appl. Phys. Lett., **96**, 021108 (2010)

[37] F. Wang, V. Pistore, M. Riesch, H. Nong, P.B. Vigneron, R. Colombelli, O. Parillaud, J. Mangeney, J. Tignon, C. Jirauschek and S. S. Dhillon, Light: Sci. Appl. **9**, 51 (2020).

[38] G. D. Rovera, and O. Acef, Optical frequency measurements relying on a mid-infrared frequency standard, A. N. Luiten (Ed.): Frequency Measurement and Control, Topics Appl. Phys. **79**, 249–272 (2001).

[39] M.Piccardo, B.Schwarz, D.Kazakov, M.Beiser, N.Opačak, Y.Wang, S. Jha, J. Hillbrand, M. Tamagnone, W. T. Chen, A. Y. Zhu, L. L. Columbo, A. Belyanin, and F. Capasso, Nature **582**, 360–364 (2020)

[40] A. E. Siegman in Lasers 313-323 (University Science Books, 1986)

[41] G. P. Agrawal, Opt. Lett. 12, 260-262 (1987)